# CONTROL OF MULTISCALE SYSTEMS WITH CONSTRAINTS.
# 1. BASIC PRINCIPLES OF THE CONCEPT OF EVOLUTION OF SYSTEMS WITH VARYING CONSTRAINTS

**S. Adamenko [1], V. Bolotov [2], V. Novikov [3]**

[1] Electrodynamics Laboratory *Proton-21*
[2] V. Karazin National University of Kharkov
[3] The Institute of Electrophysics and Radiation Technologies of the National Academy of Sciences of Ukraine

*Abstract*. Physical fundamentals of the self-organizing theory for the system with varying constraints are considered. A variation principle, specifically the principle of dynamic harmonization as a generalization of the Gauss-Hertz principle for the systems with varying internal structure is formulated. In compliance with this principle the system evolves through dynamics of the processes leading to harmonization of the internal multiscale structure of the system and its connections with external actions as a result of minimizing the dynamic harmonization function. Main principles of the 'shell' model of self-organization under the action of the dominating entropic disturbance are formulated.

CONTENTS

**1. Introduction**

People for many centuries have been concerned with the problems of predictability and predeterminacy of events or, in other more general terms, with the problems of irreversible evolution and invention of methods to influence evolution in a desirable way. Revolutionary steps in understanding the processes of evolution and self-organization of the systems of various types were made in the second half of the previous century. I. Prigogine and his school greatly contributed in resolving these problems. The interest to these problems has quickened in the recent times.

For a long time science has mainly focused on analysis of the Nature phenomena, however now is the time when the problems of synthesis and control become especially urgent. Solution of the mankind energy problems is related to solution of the problem on actual control for the nuclear structure synthesis. However, despite of successes gained in the evolution theory, still there is a lack of clear understanding of general laws of control for synthesis of new structures and evolution trends, neither there are considerable advances in synthesizing nuclear structures. Problems of controlling evolution trends have not yet been solved and it is still impossible to reliably predict consequences of technogeous interventions in the Nature evolution.

At the same time ongoing successful synthesis of new biological structures and development of technologies for creation of the gene-modified objects without clear understanding of their remote consequences for the biological evolution enhances the potential environmental threats.

I. Prigogine's works made it clear that in certain conditions under rapid growth of electromagnetic pollutions our influence on the evolution may be only inconsiderable, thus generating concerns with unpredictability of electromagnetic smog as an unconscious stimulus for the biological evolution with unpredictable consequences. There is a long-felt need to address the problem of controlling self-organization in various systems consisting of various elements.

One of the most general definitions of the system as a set of interrelated components proposed by L. Bertalanffy in the general systems theory already contains a notion of interplay between the elements, i.e.

connections. Constraints or conditions disallowing the systems elements to occupy arbitrary positions or have arbitrary velocities or other characteristics are called constraints. These are the constraints that ensure the system's wholeness, structure and stability. Evolution of the system presents, in its turn, evolution of the systems' inner structure, i.e. evolution of the system's constraints.

Unfortunately, the current self-organization theory does not use either generalized structure parameters or bonding energy and mass defects as basic variables. To our opinion, this factor is the limitation of the available dynamic systems evolution theory, which does not allow its effective application for governing the synthesis processes. Understanding of self-organization in any system as a purposeful process of varying structure and bonding energy as well as formulation of a general variation principle governing the open systems evolution serve the grounds for the theory we are developing.

Recognizing specifics in governing evolution of various systems we see common ways for solution of these problems in various spheres of knowledge from cosmology to nuclear physics and from biology to sociology.

We are offering for your attention a series of works, which state the general concept of evolution of complex systems with varying constraints and show some applications of this concept for creating new methods of governing synthesis of new structures in these systems.

The works [1-2] report the prehistory of creating the concept of collective controlled synthesis of new nuclear structures with definite energy directiveness and provide its fundamental principles based on the use of the following:
- Notions of binding energy and mass defect;
- Notions of mass forces and accelerations;
- Relations of the sign of mass defect and accelerations;
- Collective processes for effective initiation and control of the synthesis of systems with the required mass defect;
- Variation principle of the complex systems evolution – the principle of dynamic harmonization.

A notion of dissipative structures introduced by Prigogine (see, for example [3]) proved very useful in many spheres of science and engineering. Essential aspects of the dynamic systems control theory are developed in the framework of the theory of chaos, fractal physics and synergetics with their successful application in many spheres of science and technology. Nuclear physics is practically the only science where the notions of the theory of dissipative structures, dynamic chaos and self-organization are not yet effectively applicable, while solution of the problems on nucleosynthesis control is vital for obtaining answers to the following questions:
- How to solve the problem of the world's growing energy needs in the most harmonious and environmentally friendly way?
- What key phenomena and effects can serve as a basis for creating new efficient energy technologies and governing the potential energy opportunities in the surrounding Nature?

On the other hand, development of the theory of complex systems self-organization is hindered by the fact that such physical values as bonding energy and mass defect defining energy potential in any physical processes and being widely used in the nuclear physics are not yet employed in the theory of evolution.

We will try to include these new aspects into the complex systems self-organization theory in these series of works. We will provide in this work physical substantiation of the concept basic principles while theoretical development of these concepts will be reported in the next two works.

**2. Main notions of the concept: bonding energy, structure, mass defect, scaling, fractal dimension and clusters**

All primary energy sources in Nature have common basis, namely the processes of the system bonding energy variation. This is related both to the most common energy sources based on transformation of the bonding energy at the atomic and molecular levels, for example in organic fuel combustion, and in nuclear processes generated by the variation of the bonding energy of nucleons in the atom nuclei during nuclear reactions.

Under the bonding energy $B$ of a system consisting of $i$ components (particles) one understands the difference between the apparent energy of a system of bodies or particles $W$ and the total energy of the same bodies or particles in the state of equilibrium in the absence of any interaction $\sum_i W_i$:

$$B = \sum_i W_i - W, \tag{1}$$

where $W_i$ - apparent energy of $i$ component in the unbound rest state. Mass defect of the system $\Delta m$ is the difference between the sum of the system elements masses $m_i$ and total mass of the system $m$

$$\Delta m = \sum_i m_i - m. \tag{2}$$

Mass defect of the system is characterized by its stability. In addition to mass, the system inertia in response to the forces acting on the system is its another most significant characteristic. Mass of the system depends on its structure and relations within it, and determines inertia of the system at rest. Inertia in contrast to the mass depends on the coordinate system (see, for example [4]).

With all versatility of particles and systems motions, the latter may be classified. Nature abounds with periodic regular phenomena ranging from pendulum motion to atomic oscillations. Most of the real phenomena are non-linear and instead of periodicity produce aperiodic and chaotic motions while the generated geometric structures are not continuous. With all variety of nonlinear evolution there are general properties uniting many of them, which are self-similarity and invariance in relation to the scale variations (scaling). Scaling manifests itself in many non-linear physical processes, especially when studying critical phenomena characterizing behavior of the substances in the vicinity of the phase transition points.

Proceeding from the general definition of the system it follows at once that the system has at least two spatial scales – internal microscopic scale $L$, determining its specific dimensions as an integral object, and minimal spatial scale $l_1$, related to specific dimensions of minimal system elements (monomers) being the parts of the system. In this case monomers are considered as objects without internal structure. Thus, the most important characteristics of the system for general analysis of self-organization are as follows:
- Space limitation and, thus, its space scale;
- Internal structure of its relations and complexity, that may be appropriately characterized by any of the fractal dimensions;
- Space and mass characteristics of the structural elements of the system (monomers).

Clusters of various scales are a general model of such systems [5]. One may say that the systems evolves through formation of mesoscopic structures, which are clusters with scales $l_i$ satisfying the inequality $l_1 \ll l_i \ll L$.

Growth of the structure from a $A$ set of monomers naturally divides the system into two parts – the structured one consisting of $A_{cog}$ monomers and the strctureless one consisting of the remaining $A_g = A - A_{cog}$ monomers. A proportion of all monomers in composition of the structured part is called 'order parameter' $\eta \approx A_{cog}/(A_{cog} + A_g)$.

Thus, the system evolution results in formation of the cluster consisting of monomers. This cluster is composed of elements, which in their turn are fractal clusters of a smaller scale $l_i$ from $A_i$ monomers. Multiscale systems possessing scale invariance in the sphere of $l_1 \ll l_i \ll L$ scales are the most important for implementing control processes.

In the general case, constraints in the system and their complexity may be characterized by a fractural dimensions, for example, connectedness dimension $D_c$, determined by the structure of constraints or mass fractal dimension of the system $D_f$, determined by distribution of the substance in the system [6]. Dependence of the angle-averaged correlation function on a distance from its geometrical center is one of the exponential functions typical for the fractal. In this case average density of particles in the cluster $\rho(r)$ with moving away from the center within the cluster varies according to the law complying with the law of space correlations decay:

$$\rho(r) = \rho_m \left(l_m/r\right)^{3-D_f}, \tag{3}$$

$\rho_m$ - density of monomers from which the fractal structure is composed.

From these general correlations of the fractal geometry a simple correlation between the mass number of the fractal cluster $A$ (i.e., the number of monomers of which the cluster consists), the cluster overall dimension $R_A$ and the monomers characteristic dimension $l_m$ follows:

$$A \propto (R_A/l_m)^{D_f} \text{ или } R_A = l_m A^{1/D_f}. \qquad (4)$$

From the correlation (3) it also follows a dependence of the fractal dimension of the cluster with the mass number $A$ and its average density $\rho$ on the mass number $A_m$ and density $\rho_m$ of the structureless units – monomers of which the cluster is composed:

$$D_f = 3\frac{\ln(A/A_m)}{\ln(A/A_m) + \ln(\rho_m/\rho)}, \quad \rho = \rho_m \left(\frac{A_m}{A}\right)^{\frac{3-D_f}{D_f}}. \qquad (5)$$

The fractal dimension characterizes properties of the system's scale invariance related to the system's coherence parameter. Let us estimate correlation between the coherence parameter and the fractal dimension.

The coherent part of the system in the sufficiently general case may be considered as a fractal cluster. Since the potential energy of the substance is mainly proportional to the density, then it obeys the exponential law (3), true for the cluster substance. That is, the cluster potential energy is the function with the similarity coefficient $k_{sc} = D_f - 3$: $U(r) \propto \rho(r) \propto r^{k_{sc}}$, i.e. $U(\alpha r) = \alpha^{k_{sc}} U(r)$. From the virial theorem for the systems with potential energy possessing similarity a correlation between the mean values of the kinetic $\overline{W_{kin}}$ and potential energies $\overline{U}$ satisfies the equality $\frac{\overline{W_{kin}}}{\overline{U}} = \frac{k_{sc}}{2}$. If it is remembered that the coherent part possesses mainly the potential energy, while all kinetic energy is concentrated in the non-regular and non-coherent component, then one may obtain the estimate of the order parameter $\eta \approx \overline{U}/(\overline{W_{kin}} + \overline{U}) = 1/\left(1 + \frac{k_{sc}}{2}\right)$. Whence it follows:

$$\eta \approx (3 - D_f)/(D_f - 1), \quad 0 \leq \eta \leq 1. \qquad (6)$$

Each stage of the scales hierarchy ranging from the largest to the smallest scale may have its own order parameters. If the order parameters at different scales proved to be connected, then one may state that a whole multi-scale macroscopic object appeared.

## 3. Bonding energy in a system of particles with internal cluster structure

Let us consider a simple but at the same time important example of the classical multi-scale systems – a system of many correlating particles capable to transform the cluster structure inside the system. The simplest general model of such systems may be represented by a drop of liquid composed of $A$ molecules able for form clusters of several molecules.

We will denote these clusters $A_m$, where $m$ - the number of molecules in the cluster. These clusters in their turn may create the dendrite structure with the fractal dimension $D_f$. It means that the drop itself is a multi-scale macroscopic object composed of the coherent part formed by the dendrite of monomers $A_m$, comprising $\eta A$ molecules, and structureless liquid part of $(1-\eta)A$ molecules.

Correlation of the liquid molecules may be approximately described by the Lennard-Jones potential

$$U(r) = \varepsilon_0 \left((r_{min}/r)^{12} - 2(r_{min}/r)^6\right), \qquad (7)$$

which corresponds to the molecules attraction at sufficiently large distances $r \gg r_{min}$ and their sharp repulsion at smaller distances $r \ll r_{min}$. Such correlation pattern provides integrity to the system of molecules with typical distances between them $r_{min}$ and positive energy of bonds between the molecules of the order $\varepsilon_0$.

Full volumetric bonding energy $B_v$ of the liquid drop is determined by the integral $B_v = \int U(r-r')\rho(r)\rho(r')d^3r d^3r'$ and because of sharp repulsion at smaller distances it proves proportional not to the square but to the first order of the molecules quantity in the drop $A$. Corrections grow along with density because the system of particles is not ideal [7] and it may be written as:

$$B_v \approx g_0 A + a_0 \rho^{2/3} A, \quad g_0 \approx \varepsilon_0. \tag{8}$$

According to the principles of a simple and effective classical Ya. Frenkel theory of the liquid drop [8] in addition to the indicated positive contribution to the bonding energy there is a negative contribution from the surface energy of the interface boundary $B_{surf}$, which is proportional to the boundary square $S_{drop}$:

$$B_{surf} = -\sigma S_{drop}, \quad S_{drop} = 4\pi R_A^2 (A/A_m)^{2/D_f - 2/3}, \quad R_A = l_m \left(\frac{A}{A_m}\right)^{1/3}. \tag{9}$$

The system of particles in addition to the potential energy has a kinetic energy of the chaotic motion $B_{kin}$ with temperature $T$, reducing the bonding energy of the cluster $B_{kin} = -\frac{3}{2}T\left(\frac{A}{A_m}\right)$. Using for estimation a general polytropic process $\frac{T}{\rho^{\gamma-1}} = const$ with an indicator $\gamma$:

$$B_{kin} = -a_T \left(\frac{\rho}{\rho_m}\right)^{\gamma-1} A. \tag{10}$$

If a part of the system is ionized and has the charge $Z$, then instead of one component of the system with the density $\rho$, there appear three ones – neutral with density $\rho_0$, a positively charged component with density $\rho_Z$ and electron component with density $\rho_{el}$: $\rho_0 = \frac{A-Z}{A}\rho$, $\rho_Z = \frac{Z}{A}\rho$.

Contributions to the volumetric part of the bonding energy and to the kinetic energy of the neutral and charged component take the form:

$$B_v \approx g_0 A + a_0 \rho_0^{2/3}(A-Z) + a_0 \rho_Z^{2/3} Z = g_0 A + a_0 \rho_m^{2/3}\left(\left(1-\frac{Z}{A}\right)^{5/3} + \left(\frac{Z}{A}\right)^{5/3}\right)\left(\frac{A_m}{A}\right)^{\frac{2(3-D_f)}{3 D_f}} A \tag{11}$$

$$B_{kin} = -a_T \left(\frac{\rho}{\rho_m}\right)^{\gamma-1} A = -a_T \left(\frac{\rho_0}{\rho_m}\right)^{\gamma-1}(A-Z) - a_T\left(\frac{\rho_Z}{\rho_m}\right)^{\gamma-1} Z = -a_T \rho_m^{\gamma-1}\left(\left(1-\frac{Z}{A}\right)^\gamma + \left(\frac{Z}{A}\right)^\gamma\right)\left(\frac{A_m}{A}\right)^{(\gamma-1)\frac{3-D_f}{D_f}} A.$$

Moreover, new contributions appear in the bonding energy – bonding energy resulted from the Coulomb repulsion of the like-charged particles $B_q$ and bonding energy of electrons $B_{el}$ (calculated in approximation of the degenerate electron Fermi – liquid with regard of quasi-neutrality):

$$B_q = -\frac{3}{5}\frac{e^2 Z(Z-1)}{R_A (A/A_m)^{1/D_f - 1/3}}, \quad B_{el} = -\frac{3}{4} a_\varepsilon \left(\frac{\rho}{\rho_m}\right)^{4/3}\left(\frac{Z}{A}\right)^{4/3}. \tag{12}$$

By introducing a variable $y = \frac{1}{2} - \frac{Z}{A}$, one may take advantage of a convenient approximation:

$$\left(1-\frac{Z}{A}\right)^\gamma + \left(\frac{Z}{A}\right)^\gamma = \left(\frac{1}{2}+y\right)^\gamma + \left(\frac{1}{2}-y\right)^\gamma \underset{y\ll 1}{\approx} 2^{1-\gamma} + \gamma(\gamma-1)y^2 = 2^{1-\gamma} + \frac{\gamma(\gamma-1)}{4}\left(1-2\frac{Z}{A}\right)^2$$

and present the full bonding energy as:

$$B_{drop} = \left(c_0 - c_3\left(1-\frac{2Z}{A}\right)^2\right) A - c_1 A^{2/3} - c_2 \frac{Z^2}{A^{1/3}} - c_{el}\left(\frac{Z}{A}\right)^{4/3}, \tag{13}$$

$$c_0 = g_0 + a_0 \rho_m^{2/3} \frac{1}{2^{2/3}}\left(\frac{A_m}{A}\right)^{\frac{2(3-D_f)}{3 D_f}}, \quad c_1 = 4\pi\sigma l_m^2 \left(\frac{A}{A_m}\right)^{\frac{2}{D_f}}, \quad c_2 = \frac{3}{5}\frac{e^2 Z^2}{l_m}\left(\frac{A_m}{A}\right)^{\frac{1}{D_f}},$$

$$c_3 = -\frac{5}{2^{1/3} 9} a_0 \rho_m^{2/3}\left(\frac{A_m}{A}\right)^{\frac{2(3-D_f)}{3 D_f}}, \quad c_T = \rho_m\left(\frac{A_m}{A}\right)^{\frac{3-D_f}{D_f}} T, \quad c_{el} = \frac{3}{4} a_\varepsilon \left(\rho_m^{4/3}\left(\frac{A_m}{A}\right)^{\frac{4\,3-D_f}{3\,D_f}}\right).$$

Expression for the bonding energy of the drop with the cluster structure (13) allows from the equilibrium condition $\frac{\partial}{\partial Z} B_{drop} = 0$ determining the charge of the quasi-stationary cluster with a maximal bonding energy.

This algebraic equation has an exact solution, which may be approximately presented in a simple form:

$$\frac{Z}{A} \approx \frac{1}{2 + \frac{c_2}{2c_3} A^{2/3}} \quad (14)$$

Evolution of the cluster presents transitions between quasi-stationary states that may be described by the equations of the variation principle of the dynamic harmonization obtained in the end of the work.

### 4. Mass defect, production of entropy and entropic forces

One of the main problems of the evolution of systems with varying constraints is the problem on the general laws governing variations of the fractal dimension, the order parameter and the system's inner structure over time.

Variation of the fractal dimension $D_f$ is accompanied, according to (13), with variation of the bonding energy and mass defect $\delta m_i(D_f) = B_i(D_f)/c^2$. Variation of the mass defect is related to the mass force and corresponding acceleration:

$$F_m = -\frac{\Delta \delta m}{\tau} u_i, \quad a_m = -\frac{1}{m}\frac{d\delta m}{dt} u_i \approx \bar{\sigma}_S u_i . \quad (14)$$

In the last correlation it is considered that relation of the mass defect to the total mass presents the value approximately equal to the order parameter, while

$$-\frac{1}{m}\frac{d\delta m}{dt} \approx \frac{\delta m}{m}\frac{d}{dt}(-\ln \eta) \approx \bar{\sigma}_S , \quad (15)$$

where $\bar{\sigma}_S$ - average production of the entropy in the system.

Processes of the entropy production and the entropy flows are caused by the entropy forces. Apparently, existence of the entropy flows is conditioned by the entropy gradients and we may determine these forces as follows:

$$F_S = w(\eta)\nabla S . \quad (16)$$

Coefficient $w$, depending on the current order parameter $\eta$, represents the energy density of the processes related to the entropy flows. In case of the local equilibrium, $w \underset{\eta \to 0}{\to} T$ and (16) is in agreement with the expression for the mass entropy forces following from the main correlations of the locally equilibrium thermodynamics.

It is noteworthy that the entropic forces introduced in the work [9], essentially differ from the multi-scale structure-forming entropic forces (16). In (16) the entropy gradient creates the force acting through the forces of the system equally on all its particles, which is the mass force by definition.

Behaviour of the system near the phase transition presents a simple example demonstrating the appearance of the mass force. If to place a new phase nucleus into the supercooled liquid, this will generate the explosive transition to the nucleus phase accompanied by the mass entropy forces. In this case, in contrast to the intermolecular forces, this force is not directly connected with direct correlation between particles, but has a collective nature: as a result of dynamics the systems evolves through temperature-related 'trials and errors' tending to transfer from the less probable state to the more probable one.

Since the dissipative factors may be neglected in the processes of multi-scale self-organization under study while energy variations in the system are mainly related to the evolution of the bonding energy in the system under the action of the structure-forming entropic forces, then the arising structures may be called entropic (informational) in contrast to the Prigogine's dissipative structures. It is notable that thereby the system creates a memory on the action of mass entropic forces and even after termination of this action they leave traces as the formed entropic structures whose further dynamics may be determined by the Prigogine's irreversible thermodynamics.

### 5. Mass forces and flow in the phase space

Approximation of the local equilibrium is based on the assumption that the distribution functions in the variable point of the space in physically infinitesimal volumes have an equilibrium form corresponding to the assumption on the detailed balance with the flow in the phase space equal to 0. In this case the distribution function parameters (temperature, density, the Fermi energy) may depend on the point, while their evolution fully determines the evolution of the disequilibrium states. However, with sufficiently powerful forces and drains corresponding to the flows that exceed the dissipative flows in the system an essentially different physical situation may develop.

The idea about the system's elements – particles as material points causes a number of problems. For example, the electromagnetic field energy is infinite for charged point particles.

Consistent use of the system particles presentation through their distribution in the coordinate space and other kinematic variables (velocities and accelerations) remove these problems [10-12], while the main equations for the system in the collective states are not only dynamic equations of the motion of correlating particles but also kinetic equations for distribution functions of the system's elements. In the next article we will obtain and analyze the kinetic equations as generalization of the Vlasov's kinetic equation for the open systems with varying constraints while here we only give quality observations.

The kinetic equation for the system of particles in absence of external forces may be written as the continuity equation in the phase space -

$$\frac{\partial f}{\partial t} + div_r(\vec{j}) = I_{st}, \quad I_{st} = -div_p(\vec{j}), \quad j_i = D_{ij}\frac{\partial f}{\partial p_j} + F_i f, \quad (17)$$

since for the collision integral $I_{st}$ we use representation as the flow divergence in the phase space $j_i$ and the distribution function dynamics is represented by the dynamics of the effective incompressible liquid. The flow is written through the diffusion coefficient $D_{ij}$ and friction force $F$ in the phase space, which is true for the kinetic equations, which in various forms consider collective correlation of the particles through natural oscillations of the medium (equations of Vlasov, Landau or Lennard-Balesku).

Let us assume that evolution is a number of bifurcations between a sequence of quasi-stationary states, which, thus, may be considered as the ones determined by a system of equations:

$$div_r(\vec{j}) = 0; \quad div_p(\vec{j}) = 0. \quad (18)$$

The first equation is apparently satisfied in spatially homogeneous systems and the stationary distribution functions for them should satisfy the equation:

$$div_p(\vec{j}) = 0 \text{ или } |\vec{j}| = P, \quad P = const. \quad (19)$$

Solution of this equation with permanent flow $P$ (see. [13-14]), different from zero, corresponds to the action of the mass forces on the system.

Genuinely, the system by definition is called 'mass' if it acts not only on the particles on the boundary of the system, but also on all particles inside the system. If in this case the mass force exceeds the dissipative forces in the system, then dissipation inside the system may be neglected, while all points inside the systems may be considered as approximately equivalent. The mass forces exceeding the dissipative ones will be called the general dominating disturbance for this system. Therefore, the system under the action of the general dominating disturbance may be well simulated by the spatially homogeneous non-equilibrium system with flows in the phase space constant in each point of the space.

Vlasov analyzed in [10-12] and in [15-16] important examples of physical mechanisms fostering formation of structures in plasma due to renormalization of particles correlation through natural collective oscillations of the medium.

The renormalization of particles correlation is determined by their distribution function. It is generally regarded that the only stationary solution of the kinetic equations in the spatially homogeneous system is the equilibrium distribution function $f_T(p) \propto \exp(-p^2/(2mT))$, corresponding to $P = 0$ and representing the trivial solution (17). It appears to be that the stationary states of the spatially homogeneous systems in case of permanent, not equal to zero, flows in the phase space have a power form or power asymptotics (see [13-14]).

As it will be shown in the next work, it is convenient to represent the power solutions in the form of the solution of the generalized Vlasov's equation:

$$f_q(\varepsilon) = A\, exp_q(-\varepsilon/T), \quad exp_q(x) = (1+(q-1)x)^{1/(q-1)}. \quad (20)$$

Here for notation of solutions we use the quasi-power generalizations of the exponential functions introduced by C. Tsallis for his open systems thermodynamics [17]. The distribution functions (20) describe the collective state resulted from the action of the mass force while the parameter $q \approx 1 + \alpha P$ is determined by the value of the corresponding flow $P$. Average energy on the particle in the non-equilibrium state (18) is equal to:

$$T(1+(1-q)S_q), \quad (21)$$

where $S_q = -\ln_q \eta$ – the entropy of the state (18), written with the use of the expression for the generalized algorithm $\ln_q(x) = \dfrac{x^{1-q}-1}{1-q}$ [17]. Solutions of the (20) type in absence of the flow (i.e. with $P = 0$ and $q = 1$) form the homogeneous equilibrium state.

In the general case the system has two components:
- A structureless one with thermal motion and, evidently, distribution approaching equilibrium;
- A coherent one related to the appearance of the long-range orders and with (18) type distribution.

Since proportion of the number of particles in the components is regulated by the order parameter, then it is possible to use further and with good accuracy the model representation for the distribution function as follows:

$$f(p,T,q) = (1-\eta) f_T(p) + \eta f_q(p). \tag{22}$$

It is noteworthy that existence of the two components of the distribution function with different average energy leads to the appearance of a new branch of the medium natural oscillations with linear dispersion [18] and to renormalization of the Coulomb interaction in compliance with the theory [19], owing to the interaction through the exchange of these quantums of the medium natural oscillations. Renormalization depends on the order parameter and $q$, i.e. finally on the mass forces and flows in the system.

## 6. Shell model of self-organization of a system with constraints

Gravitational forces are the most well known examples of the mass forces because being in the cosmic scale the main factor of evolution, they lead, as the experiments show, to formation of the specific structures in the evolving Universe – the 'pancakes' that further evolve into a cluster of galaxies. Increase of the system coherence and decrease of characteristic dimensions towards coherence due to the action of the mass forces are, to our opinion, one of the key elements for initiation of self-organization in any complex systems.

In the general case, a system, isotropic at the initial moment and with distribution of particles in the space with a typical scale $l_0$ evolves into a deformed state with the large number of external spatial scales. Let us in the simplest case consider evolution anisotropy by introducing two scales as macroscopic geometric characteristics of the system instead of only one – its radius:
- $l_\perp < l_0$ in one direction;
- $l_s > l_0$ in orthogonal directions.

A smaller scale may be called the scale of the space coherence of the system, which characterizes the 'pancake' thickness, while the larger scale – the characteristic scale of interaction, which characterizes the maximal size of correlations in the system.

The phase volume of the system $\Omega_{ph} = \Omega_p \Omega_r$ is the product of the volumes. Here $\Omega_r$ is the volume in the coordination space, while $\Omega_p$ - in the pulse space. The phase volume of the system may be estimated by the distribution function, hence the quasi-stationary distribution function of the system $\dfrac{\partial f}{\partial t} \approx 0$ is followed by the stationary condition of the phase volume $\dfrac{\partial \Omega_{ph}}{\partial t} \approx 0$ under meeting the condition of the flow constancy (19). If the flow equals zero, the state of the system is in equilibrium, while $\Omega_{ph}$ is constant.

Methods of the regularization on the basis of the fractional operators developed in the work [20] may provide qualitative description of the properties of the quasi-stationary states with the non-zero constant flow in the phase space. Stationary condition of the phase volume regarding regularization takes the form:

$$D^{1-\nu} \Omega_{ph}(t) = 0, \quad (D^\alpha f)(x) = -\frac{1}{\Gamma(1-\alpha)} \frac{d}{dx} \int_x^\tau \frac{f(t)}{(t-x)^\alpha} dt \tag{23}$$

and has a solution

$$\Omega_{ph}(t) = \Omega_{ph0} \frac{\tau^\nu}{(\tau - t)^\nu}, \tag{24}$$

where $\Omega_{ph0}$ is the value of the phase volume at the initial moment of time. The derivative index $\nu$ in this correlation is proportional to the mass force or flow in the phase space and transforms into an equilibrium expression at $\nu = 0$. The above obtained expression for the phase volume of the system under the action of the mass forces shows the appearance of the mode with aggravation and reduction of the coherence scale $l_\perp$ and leads to an increase of the typical interaction scale $l_s$ in compliance with the correlation:

$$l_s = l_0 \sqrt{\frac{\Omega_{p0}}{\Omega_{ps}}} \sqrt{\frac{l_0}{l_\perp}} = l_0 \frac{1}{(1-t/\tau)^{\nu/2}} \sqrt{\frac{l_0}{l_\perp}}. \tag{25}$$

For the typical interaction scale in the correlation (25) the phase volume of the systems is not preserved while an additional rheonomous multiplier appears and explosively changes the localization of the states due to the entropy production.

To our opinion, initiation of the evolution not only in the cosmological domain but also in the general case is related to the following: increase of the system coherence is accompanied by the renormalization of the fundamental interactions in the system and corresponding changes in its structure through the increase of the typical space scales of interaction in the directions orthogonal to the direction of the coherence. An effective reduction of the dimensionality of many particles in the direction of the coherence growth of the system and 'flattening' of its collective state take place.

There appears a new class of phenomena related to the quantum nonlocality while appearance of the coherent states and processes governed by the external sources of energy and information act as a prerequisite.

In the course of interaction the quantum systems acquire classical features, which correspond to the information available in the external sources affecting the quantum system while nonlocality appears as a result of entanglement of the quantum states in the irreversible processes under interaction with the medium. One may say that the evolution of the open nonlocal quantum system generates internal information-intensive structures capable to exchange information with external sources. In this situation the system loses many of its specific quantum features and becomes to a certain degree classical. The theory of such macroscopic quantum objects and quasi-stationary states will be reported in the next articles on the basis of the modified Vlasov's equation.

Self-consistent full ionization of substance is the physical basis of the modes with aggregation. Ionization is known to lead to a density increase when electron shells of atoms reduce their radius in the process of ionization. Increase of the substance density may induce further ionization of the substance, which does not take place under the normal conditions because density increase resulted from ionization is not enough for further ionization, thus providing stability of the surrounding substance preventing its spontaneous collapse.

However, it is possible to create collapse of the electronic system by using renormalization of the electromagnetic interaction in the medium thanks to its polarization, which fits perfectly into the formalism of the dielectric permeability $\varepsilon(\omega, \vec{k})$ and exchange of quanta of the medium natural oscillations.

Fractal properties of the medium related with its scale invariance, resonance properties and space limitation of the particles subsystems are the most essential factors, which permit controlling properties of the dielectric permeability (and thus, correlation of ions and nuclei). That is, the initial Coulomb interaction of nuclei with the Fourier-transform of the potential written in the form $U(\omega,k) = \frac{4\pi Z^2 e^2}{k^2}$ essentially changes as a result of interaction through collective plasma oscillations and is determined by the dielectric permeability $\varepsilon(\omega, \vec{k})$ of the medium: $U(\omega,k) = \frac{4\pi Z^2 e^2}{\varepsilon(\omega, \vec{k}) k^2}$.

We will show that presence of the fractal structure in the system leads to the renormalization of the vacuum interaction and further spontaneous growth of fractal structures in its volume.

Fractals with the Cantor set structure (see [21]), which are built by a similitude of the bounded interval with eliminated central part whose size equals to $\xi^{th}$ fraction of the whole interval $0 < \xi < 0.5$ are convenient for simulation and theoretical studies.

Let us consider a thin fractal layer of the length $L$, which the Cantor set with the preset parameter $\xi$ and, thus, with fractal dimension $D_f = \frac{\ln 2}{|\ln \xi|}$. Distribution of the potential and charge in such fractal thin layer is considered homogeneous through all its thickness. Distribution of the charge density in the perpendicular direction $x$ (along surface of the layer where the charge density distribution is the Cantor function $\Delta_\xi(x)$) is described by the Poisson equation:

$$\frac{d^2U}{dx^2} = -4\pi e \Delta_\xi(x) \qquad (26)$$

The Fourier component of the potential in this layer is ($k_\perp$ - the wave number along layer):

$$U_k = \frac{4\pi e}{k^2}\gamma_\xi(k_\perp L), \quad \gamma_\xi(k_\perp L) = \prod_{n=0}^{\infty}\cos\left[\frac{(1-\xi)\xi^n}{2}k_\perp L\right] \qquad (27)$$

Expression for the Fourier transform (25) coincides with the general expression through the dielectric permeability at $\varepsilon(k_\perp) = \dfrac{1}{\gamma_\xi(k_\perp L)}$.

Fig. 1 shows the Fourier component of the Coulomb potential in the medium with the fractal structures.

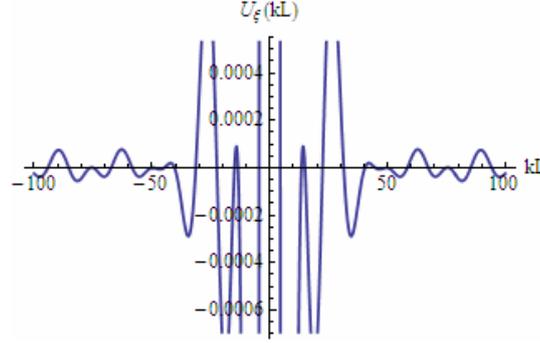

**Fig.1. Graph of the Fourier transformation of the potential in the fractal layer $\gamma_\xi(k)$, $\xi = 0.3$.**

The medium fractality leads to the appearance of the large number of the wave vector domains where the dielectric permeability is negative and interaction of the similar charges has a nature of attraction.

The inverse Fourier transformation – the component $U_k$ leads to the dependence of the potential on the coordinate for the fractal layer as the fractal function where the measure of non-zero values equals zero. That is, the volume of domains with zero values of the potential fills practically all space, while the volume of domains with the non-zero values of the potential tends to zero.

A set of points, on which the values of potential are concentrated, forms the Cantor set while the potential may be represented as follows:

$$U(r) = \sum_{i \in \Delta_\xi(r)} \frac{\exp(-\kappa_r r_i)}{r_i}, \quad \kappa_r = \frac{2\pi}{\delta_{sh}}, \qquad (28)$$

where $\Delta_\xi(r)$ - the Cantor set on which the values of the potential are concentrated, $k_r$ - the wave number of the wave along direction $r$, where the shell is limited and coherent, $\delta_{sh}$ - thickness of the shell.

Since almost everywhere (except a set of points of the zero measure) the Coulomb interaction proved suppressed and nothing prevents the particles to contact, then the fractal structure initiates its explosive growth.

A system of particles aggregated as a result of the pair contacts represents a set of clusters of various sizes. Size distribution of the clusters, i.e. concentration of the clusters of $k$ size (clusters composed of $k$ nucleons) as a function of time is described by a system of reactions:

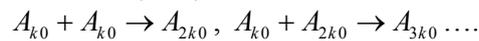

In this case the equation for concentrations $C_k$ of clusters of $k$ nucleons may be described as the Smolukhovsky's coagulation equation [22], where competition of two processes is considered: (1) adhesion of the cluster components, i.e. increase of the cluster size, and (2) collapse of the cluster components, i.e. growth of the number of clusters smaller by weight. For probability $K_{ij}$ of the adhesion of clusters of the sizes $i$ and $j$ one may assume approximation under which this probability is proportional to the product of the areas of the initial clusters surface – $K_{ij} \propto (ij)^{2/3}$.

The Smolukhovsky's equation may be integrated analytically in this approximation of the permeability of the Coulomb nuclei barriers due to the increase of correlation in the system, and it appears that the average size of the cluster may become infinite for the finite time – the time of phase transition in gel.

Solution of the problem on determination of the most general laws of the structure growth is the most important element of the self organization theory, which is reported below.

## 7. Variational principle for evolution of complex systems – a principle of dynamic harmonization in the non-covariant Gaussian form

Variational principles are the most concentrated expression of the laws on dynamics of the particles system, therefore, it is desirable to formulate the laws of the complex systems evolution in terms of the variational principles.

The variational principles of mechanics are its fundamental principles expressed in the form of variational correlations from which differential equations of motion logically follow.

According to the variational principles, actual motions of the system under the action of the preset forces are compared with the kinetically possible motions prompted by the constraints applied on the system and satisfying certain conditions. The variational principles differ by their form, variational ranges as well as by generality, however each principle incorporates everything in this sphere of science and unites all its principles in one formulation in the frames of its applicability.

In [2] the variational principle of evolution of the systems with constraints was formulated, which is the principle of dynamic harmonization. The system self-organization results from variation of the structure of constraints between the elements of the system in response to the system acceleration, and is aimed at:

- either counteraction to the forced acceleration because of the steady state due to the system energy inertia;
- or facilitation of the forced motion acceleration towards steady state at the account of the system inertia decrease.

Changes in the system structure lead either to binding of free energy of the external accelerating mass force in the structure or to a release of the previously conserved free energy in various forms into the environment.

Below we will explain this principle and its analytical formulations.

The variational principles may have various forms for the dynamics of the mechanical systems different in quality.

The systems with constraints may be open (exchanging energy and/or mass with the environment) and closed or conservative (not changing its energy and mass). Evolution of the system is always connected with structural variations, therefore with variations in the bonding energy of the evolving system and with mass defect, so in evolution processes we deal only with open systems. As to the constraints in the systems, they have more detailed classification in mechanics. Let a system be characterized with the coordinates $x^i$ and velocities $u^i = dx^i/dt$, $i = 1,...,n$. The constraints existing in the system in the general case are characterized by a set $m$ of functions:

$$\varphi_j\left(x^1....x^n, u^1....u^n, t\right) = 0, \quad j = 1,...,m. \tag{29}$$

As is know, the constraints are called:

- scleronomic, if functions $\varphi_j$ are not time-dependent;
- rheonomic, if functions $\varphi_j$ are time-dependent;
- holonomic, if functions $\varphi_j$ are not velocity-dependent;
- non-holonomic, if functions $\varphi_j$ are dependent not only on the coordinates but also on the velocities.

It is clear that constraints in the evolving system of general position will vary with time, i.e. rheonomic, but may also be holonomic and non-holonomic.

Let us change the analysis of the variation principles describing development of system particles with time.

The variation principles differ from one another by forms and varying patterns as well as by the generality degree.

The most general differential principles characterizing the motion properties of open systems with constant and variable constraints for any given point in time are the Gauss and Hertz variation principles, while the most general integral principle characterizing the motion properties at any finite time intervals, is the least action principle in the Hamilton – Ostrogradsky form [23].

For constructing the variation principle for the self-organizing open systems we start from the most general variation principle of dynamics, which is also true even for systems with non-stationary non-holonomic constraints – the Gaussian principle.

Gauss introduced the general principle of mechanics as the mechanical analog of the least square method underlain all statistical studies and it is called *the principle of least constraint*. According to the Gauss principle, positions occupied by the points of the system at the moment $t + \tau$ in true motion are distinguished among all positions allowed by the constraints by the fact that the constraint measure in them $Z_G = \sum_{i=1}^{N} m_i s_i^2$ has a minimal value (here $s_i$ – length of the vector between the points representing true or any other position of the point). The Gauss principle has the following peculiarities:
- addition of inertia mass forces to the external forces acting on the system;
- varying of accelerations under preset coordinates and velocities (the Gauss variation).

The inertia mass force proves inseparable from the corresponding acceleration, which shows its key role in self-organization processes. For the closed systems (the systems in which Hamiltonian is explicitly time-nondependent) the Gauss principle is reduced to the principle of the Hamiltonian least action.

However, even in the general variation principle of mechanics the non-stationary constraints in the system are assumed to be fully prescribed prior to the dynamic process initiation, hence this principle in its initial form cannot be the basis for the self-organization theory. Dirac [24] was the first to consider the dynamic systems with the variable structure where not only trajectories of particles but also constraints were viewed as variable parameters.

For self-organizing systems with the particle dynamics occurred in the configuration space it is necessary to take into account a possibility for the system of particles to evolve through varying the constraint fields and generation of system's most steady and optimal structures.

Moreover, since variation of the system's inner structure is connected with variation of its mass (with mass defect of the system), such processes are most effective in the course of the system evolution and may serve as a source of energy for the evolution itself. *Therefore, it is evident that control of the system with the help of the laws of evolution of its constraints (variational principles for the systems with varying constraints) is the only effective way for desirable transformations in the system at the account of its internal energy resources instead of direct 'forcing' the system by external energy only.*

It follows from the above that the use of the general dominating disturbance specially selected for the given system is the tool for initiation of the self-organization processes (dynamic harmonization) of the structure of constraints in the system.

In order to write analytically the principle of dynamic harmonization let us calculate a shift of the particle $s_i$ as a result accelerations variation. Considering time variation $\tau$ small with the accuracy to the second order it appears that:

$$s_i(t+\tau) = \frac{1}{2}\delta a_i(t)\tau^2, \text{ where } \delta a_i = w_i - \frac{F_i(t) + F_m}{m_i(D_f)}. \tag{30}$$

By inserting the shift of particles from (30) into the constraint function we receive: $\sum_{i=1}^{N} m_i s_i^2 = \sum_{i=1}^{N} \left( \frac{\tau^2}{2} (m_i w_i - F_i + (m_i a_m) u_i(t)) \right)^2 / m_i$. From here it follows that in view of the value $\tau^4/(4m_i) > 0$ the dynamic harmonization function may be represented as follows:

$$Z_{dh} = \sum_{i=1}^{N} \left( m_i(D_f) w_i - (F_i + F_m) \right)^2, \quad m_i(D_f) = \left( A_i m_p - \delta m_i(D_f) \right). \tag{31}$$

*Finally, the principle of the dynamic harmonization may be formulated as follows: the system varies its trajectory and structure under the action of external forces so that to be in the harmony with the environment and external actions as a result of minimization of the generalized constraint function $Z_{dh}$ regarding all constraints in the system. In other words, the system tends to make trajectories of its forced motion under the action of mass forces maximally approaching the trajectory of its own natural undisturbed motion.*

In (31) summation is performed both by collective variables and by all particles. Since the structure variation is inseparably connected with the variation of entropy and information, the dynamic harmonization principle simultaneously describes purposeful exchange of information and entropy by the system with the environment.

At the first glance, a property of the quadratic function minimality (31) is explicit and produces nothing new but the Newton equation $m_i w_i = F_i + F_m$. However, this is not true. From (31) *for variable determining the state of the system and its internal structure, after accelerations variation $a_i$ regarding constraints at the fixed positions and velocities of all particles follow differential equations, which do not coincide with the Newton equations for dynamic of particles under the action of forces when constraints are in place.*

In order to obtain a specific form of the dynamic harmonization equations and effectively apply it, it is necessary to use an expression for constraints in the system. It turns out that all open systems with varying constraints have significant similarities, and can suggest a general model for the evolution of such systems based on the principle of dynamic harmonization.

**8. Dynamic harmonization equation**

In accordance with the dynamic harmonization principle the evolution equations are determined by a minimum of the dynamic harmonization function $Z_{dh}$ under variation.

Let us consider an example of the liquid drop with radius $R$, which depends on the structure internal system using the equation of constraints $R = g(D_f)$, and write the dynamic harmonization function for it:

$$Z_{dh} = \frac{1}{2}(m w_R - F_R)^2 , \quad m = m_0 - B_A(D_f)/c^2. \tag{32}$$

For applying the principle let us consider that force $F_R$ may be expressed through the bonding energy gradient $B_A(D_f)$:

$$F_R(D_f) = \frac{\partial B_A(D_f)}{\partial R} \tag{33}$$

Not counting the constraints, the conditions of the constraint quadratic function lead to general Newton equations. However, owing to the constraint $R = g(D_f)$ acceleration $w_R$ cannot vary independently and is expressed through acceleration of the fractal dimension under preset values of the coordinates and velocities (the Gaussian variation of accelerations). The Gaussian variations are the variations of the second order tangent plane of the tangency at a fixed plane of the first order of tangency. Variations of the accelerations of all orders, i.e. vectors in the respective different planes are independent, therefore the Gaussian variations lead to the following: the correlations for the variations of accelerations are similar to correlations for the variation of the respective coordinates, and the first derivatives are absent in the following correlations for accelerations:

$$w_R = \frac{d^2}{dt^2} R = g_R \ddot{D}_f , \quad g_R = \frac{\partial^2 g}{\partial^2 D_f} . \tag{34}$$

Inserting the obtained expression for acceleration into $Z_{dh}$, we obtain the dynamic harmonization function as dependent on acceleration of the fractal dimension:

$$Z_{dh}(\ddot{D}_f) = \frac{1}{2}\left( g_R \ddot{D}_f - \frac{F_R(D_f)}{m} \right)^2 \tag{35}$$

Condition for minimum of the dynamic harmonization function in relation to accelerations of the fractal dimension – $\dfrac{\partial Z_{dh}(\ddot{D}_f)}{\partial \ddot{D}_f} = 0$ – leads to the differential equation determining evolution of the dynamic system with varying constraints:

$$m_{str} R_0 \ddot{D}_f - F_R(D_f) = 0, \quad m_{str} = m g_R. \tag{36}$$

In the simplest case when considering evolution of the system with slowly varying forces the equation may be once integrated and presented as the Lagrange equation with the corresponding Lagrangian function:

$$L_{stri} = m_{str}(D_f) R_0 \frac{\dot{D}_f^2}{2} + B_A(D_f), \tag{37}$$

where the system structural inertia appears $m_{str}(D_f)$. Analysis of the specific models of the system of particles regarding correlations of (25) type for space scales and bonding conditions (for example, nuclear structure models) shows an explosive growth of structural inertia with the order parameter growth, which may be approximated by the dependence $m_{str}(D_f) = \dfrac{m_{0str}}{(1-\eta(D_f))^\gamma}$. Such dependence provides hysteresis phenomenon under structure formation. Under the action of the forces with positive acceleration, structures form from the state with initial value of the order parameter while the order parameter grows and achieves corresponding maximal value. Acceleration reverses its sign while the action of the mass forces is coming to an end and these forces tend to zero, and the order parameter somewhat decreases in compliance with the harmonization equations (the Lagrangian equation with the Lagrangian function $L_{stri}$). However, since the structural inertia has already grown, the order parameter does not reduce down to its initial value and the 'residual' order parameter appears as an element of the system memory.

Structural inertia (mass) appears in the phenomena accompanied by the symmetry disturbance. This fact is well known in the theory of elementary particles. Spontaneous disturbance of the symmetry in the calibration theories may lead to the appearance of the finite mass in massless calibration particles.

After completion of the action of external entropy force, which disturbed a symmetry of the system, the long-range order, characterized by the parameter $\eta$, may appear in the system. It is the structural inertance, characterized by the order parameters at each hierarchy level that accounts for inertance of the self-organization processes. In the general case the structural inertia (mass) $m_{str}$, connected with the appearance of the fractal clusters is expressed through the fractal dimension or the order parameter.

**Conclusion**

Consistent theoretical development of the basic principles of the concept on self-organizing synthesis of new structures in the dynamic systems opens new potentials for creation of the theory on self-organization of the complex systems and for the development of fundamentally new technologies.

This work proposes generalization of the Gaussian variation principle, which is a mathematical formulation of the dynamic harmonization principle for the open systems with varying constraints. The proposed variation principle allows obtaining equations that describe the self-organization process and expose the nature of the constraints fields and their collective states.

It makes intuitive sense that self-organization of the system is inseparably connected with evolution of its structure and leads to the changes in its mass, stability and bonding energy. However, the available self-organization theories (for example, the Prigogine non-equilibrium thermodynamics) where the system self-organization is determined by the gradients of the thermodynamic parameters inside the system, while the distribution functions are locally equilibrium, are not applicable for the open systems with varying constraints. At the same time the theory on the basis of the variational principle of dynamic harmonization may claim to become the general theory on self-organization of the open systems with varying constraints.

Formalism of the variation principle of dynamic harmonization presented in the work suggests a general platform for solution of the problems of self-organization and control for evolution of various complex systems from the general positions of the theory of thermodynamic systems with varying constraints.

Since entropic forces have a considerable contribution to the proposed theory of self-organizing synthesis it is necessary to mention Kozyrev's works (see, for example [25]), where time plays a key role. Kozyrev also considered the open systems with not only degradation processes (law of the energy degradation), but also the processes of the structure synthesis and, thus the entropy reduction. He stated that the time density value, introduced in his theory, depends in a given point of the space on the processes occurring in the vicinity of this point. In the processes, where the entropy grows, the time density increases, therefore such processes are time-emitting. Hence, time density increases when a substance loses its organization. Kozyrev noted that even this circumstance suggests a conclusion that time contains organization or negative entropy that may be transmitted to another object in the vicinity of such processes. In other words, time affects the substance. It is especially interesting that Kozyrev could observe evolution in stars and stars motion from the laboratory on the Earth using his time theory in the real time mode. Using our theory on self-organization of systems with varying constraints it is possible to disclose the essence of time density variation according to Kozyrev.

In the next works we will show that this is connected with metrics change and appearance of the local space-time curvature determining in the simplest cases the value of the natural and laboratory time relations.

Herz paid attention to the fact that varying accelerations according to Gauss with minimality of the constraint function corresponds to the variations in motion on condition of the trajectory curvature minimality, hence real motion always chooses the straightest path coordinated with constraints. This principle is represented by Herz also in the form of the functional minimum denoting the length of the system path from $n$ particles with weights $m_i$.

It will be shown that the principle of dynamic harmonization regarding the Hertz ideas can be transformed to the functional minimum representing not the length of the path in the three-dimensional space but the length of world lines of particles in space - time stated and formulated as follows: *The system with constraints evolves in space-time by geodetic lines with the space - time curvature tensor corresponding to the evolution of the internal system constraints, being harmonized in response to the action of mass forces*

The works of this series will show that the idea of a liquid drop with a fractal structure can be naturally applied to a drop of nuclear liquid and a range of possible values of the nuclei binding energies is much broader than it is accepted in nuclear physics. The appearance of the internal structure of nuclei in the nucleon scale, which is reflected by introduction of a new nuclear option - their fractal dimension, opens up great prospects for synthesis of new nuclear structures based on self-organization of nuclear matter obeying the dynamic harmonization principle.

The concept of self-organizing synthesis allows obtaining theoretical and experimental results that may be applied in many fields of science and technology.